# A Universal Gauge for Thermal Conductivity of Silicon Nanowires With Different Cross Sectional Geometries

Jie Chen,[1] Gang Zhang,[2,*] and Baowen Li[1,3]

[1]Department of Physics and Centre for Computational Science and Engineering, National University of Singapore, Singapore 117542, Singapore

[2] Key Laboratory for the Physics and Chemistry of Nanodevices and Department of Electronics, Peking University, Beijing 100871, People's Republic of China

[3]NUS Graduate School for Integrative Sciences and Engineering, Singapore 117456, Singapore

*E-mail: zhanggang@pku.edu.cn

## Abstract

By using molecular dynamics simulations, we study thermal conductivity of silicon nanowires (SiNWs) with different cross sectional geometries. It is found that thermal conductivity decreases monotonically with the increase of surface-to-volume ratio (SVR). More interestingly, a simple universal linear dependence of thermal conductivity on SVR is observed for SiNWs with modest cross sectional area (larger than 20 nm$^2$), regardless of the cross sectional geometry. As a result, among different shaped SiNWs with the same cross sectional area, the one with triangular cross section has the lowest thermal conductivity. Our study provides not only a universal gauge for thermal conductivity among different cross sectional geometries, but also a designing guidance to tune thermal conductivity by geometry.

# 1. Introduction

Silicon nanowires (SiNWs) have shown tremendous promising applications in electronics, such as high-performance field-effect transistors (FETs) [1], nano sensors [2] and logic gates [3]. They have also attracted wide attention in thermoelectric applications because of their remarkable improved thermoelectric figure of merit, which is mainly caused by the significant reduction of thermal conductivity in SiNWs [4, 5]. Many approaches have been proposed to further reduce thermal conductivity of SiNWs, such as introduction of impurity scattering [6, 7], holey structure [8-10], and surface roughness [11, 12].

The well-developed fabrication technologies have offered a quite flexibility in the synthesis of nanowires with controllable diameter and cross sectional geometry [13-15]. Theoretical studies have revealed that different geometries are associated with different dynamics [16], which leads to completely different transport behavior of elastic wave in phononic crystal at macroscopic level [17]. Although theoretical [18] and experimental [19] studies have shown that thermal conductivity of SiNWs increases with the increase of transverse dimension, only fixed cross sectional geometry is considered so far. The effect of cross sectional geometry on thermal conductivity of nanostructures has not yet been explored.

In this paper, we systematically study thermal conductivity of [100] SiNWs with different cross sectional geometries, including rectangle (square), circle and triangle. For simplicity, SiNWs with rectangular, circular, and triangular cross sectional shapes are named as rect-SiNWs, cir-SiNWs, and tri-SiNWs, respectively. We found that thermal conductivity decreases linearly with the increase of surface-to-volume ratio (SVR), and this linear dependence is universal regardless of cross sectional geometries for SiNWs with modest cross sectional area (larger than about 20 nm$^2$). In contrast to the geometry sensitive heat energy transport observed in macroscopic crystal [16, 17], our results demonstrate that at nanoscale, the surface-to-volume ratio has more remarkable influence on phonon conduction than specific geometrical shape does.

## 2. Models and Numerical Results

In our study, the longitudinal direction of SiNWs is set along $x$ axis, and the transverse direction is cut into specific geometry with the cross sectional size controlled by the diameter $D$, which is defined for all geometries as the maximum distance between the surface atoms in the cross sectional plane. The maximum cross sectional area considered in this study is 806 nm$^2$ with rectangular cross section. For cir- and tri-SiNWs, there exists certain discrepancy between the actual lattice structure and ideal geometry (Fig. 1), due to the discrete stacking of Si atoms. Compared to the ideal geometry, there are more surface atoms in the actual lattice structure of cir- and tri-SiNWs. However, these additional surface atoms are not included in the conventional definition of SVR (ratio of surface area to volume) based on ideal geometry. To take into account this discrepancy, here we define SVR at atomic level as the ratio of the number of surface atoms to the total number of atoms, which can accurately describe the surface morphology.

In our non-equilibrium molecular dynamics (NEMD) simulations, Stillinger-Weber (SW) potential [20] and Langevin heat reservoir are used to calculate the thermal conductivity. SW potential has been widely used in the study of thermal and mechanical properties of silicon nanostructures [21-24]. For instance, it has been successfully used in the simulations of the crystalline structures and thermal conductivity of thin SiNWs with the diameter of 1.4 nm [21]. Moreover, it has been demonstrated that the thermodynamic property predicted by SW potential is comparable with that from Tersoff potential [25]. The velocity Verlet algorithm is employed to integrate Newton's equations of motion numerically, and each MD step is set as 0.8 fs. Fixed boundary conditions are applied to both ends of the nanowires. Multiple layers of the nanowires are attached to Langevin heat reservoir with different temperature to remove the boundary effect. More technical details about NEMD simulations can be found elsewhere [26]. Since quantum effect on thermal conductivity of SiNWs is quite small at room temperature [7], we do not adapt quantum correction in our study as we mainly concentrate on the cross sectional effect on thermal conductivity of SiNWs above room temperature.

We first calculate thermal conductivity of [100] SiNWs with different cross sectional area and geometries at fixed length $L$=3.26 nm in the longitudinal direction. In our simulations, we keep the temperature difference small and we find it has little effect on the calculated thermal conductivity. For instance, the calculated room temperature thermal conductivities of rect-SiNWs with SVR=0.083 are 6.85±0.35 W/mK, 6.56±0.26 W/mK and 6.66±0.29 W/mK under the temperature difference of 10 K, 20 K and 30 K, respectively. Fig. 2(a)-(c) shows the thermal conductivity $\kappa$ versus SVR at different temperature. At each temperature, thermal conductivity decreases monotonically with the increase of SVR for any given cross sectional geometry, which is a consequence of the enhanced surface scattering when SVR increases. More interestingly, for different cross sectional geometries, thermal conductivity follows the same linear dependence on SVR when the cross sectional area is greater than certain threshold of about 20 nm$^2$ (Fig. 2), which corresponds to a threshold value around SVR=0.14. This important feature suggests that SVR can serve as a universal gauge for thermal conductivity of SiNWs with modest cross sectional area, regardless of the specific cross sectional geometry. For very thin

SiNWs with cross section area below the threshold, the dependence of thermal conductivity on SVR deviates from the universal linear fitted line, but also depends on the specific cross sectional geometry. In the following part, we refer to the cross sectional area above and below this threshold as the "universal region" and "non-universal region", respectively.

In the universal region, the linear dependence of thermal conductivity on SVR holds for all temperature above 300 K, while the absolute value of the slope for the linear best-fit decreases with the increase of temperature. In order to understand this temperature dependence, we consider the variation of $\kappa$ of two nanowires with different SVR. When the temperature increases from room temperature, $\kappa$ of both nanowires decreases as a consequence of stronger anharmonic phonon-phonon scattering at high temperature. The higher SVR value can induce a stronger surface scattering and thus result in more localized phonon modes [8]. The introduction of phonon localization can weaken the temperature dependence of thermal conductivity [27], which has also been reported in the recent study of silicon nanotubes [8] and the crystalline-core/amorphous-shell silicon nanowires [28]. As a result, thermal conductivity of the nanowire with lower SVR decreases more than that of the one with higher SVR, leading to a decrease in the absolute value of the linear slope with the increase of temperature.

Another factor that influences the slope is the length of SiNWs. Fig. 2(d) shows the room temperature thermal conductivity versus SVR for SiNWs with length $L$=6.52 nm. Compared to Fig. 2(a), the universal linear dependence of thermal conductivity on SVR holds as well at longer length, with an increased absolute value of the slope. Thermal conductivity of low dimensional materials is quite different from that of their bulk counterparts in the sense that it is usually length dependent. For instance, Chang *et al.* experimentally discovered the length dependence of thermal conductivity in carbon and boron-nitride nanotubes at room temperature [29]. Moreover, Yang *et al.* numerically studied thermal conductivity of SiNWs with length up to 1.1 μm. They found that the length dependence of thermal conductivity persists even when the length is much longer than the phonon mean free path [30]. To understand the length dependence of the slope, we consider the variation of $\kappa$ of two nanowires with different SVR. Due to the excitation of more long-wavelength phonons which contribute a lot to thermal conductivity, κ of both nanowires increases when the nanowire length increases [30]. The higher SVR value can induce a stronger surface scattering and thus result in more localized phonon modes, which lead to the reduction of thermal conductivity [8]. Therefore, thermal conductivity of the nanowire with lower SVR increases more than that of the one with higher SVR, which gives rise to an increase in the absolute value of the linear slope when the nanowire length increases.

To further check the validity of this linear dependence at much larger dimensional scale, in the inset of Fig. 2(d), we also show the thermal conductivity versus SVR relation based on the experimental results from Ref. [19] for cir-SiNWs at 300 K. The SiNWs used in Ref. [19] are single crystalline with the length of several microns and cross section area up to $1.04 \times 10^4$ nm$^2$. A good linear dependence of thermal conductivity on SVR also appears from their results (in Ref. [19], they showed the thermal conductivity versus temperature relation for different diameters). The good agreement of experimental results with the linear fit line further supports the fact that the linear relation between thermal conductivity and SVR is an intrinsic phenomenon in NWs.

One important application of the present study is to tailor thermal conductivity by geometry. For the same cross sectional area, SVR associated with different geometries is intrinsically different, leading to different thermal conductivity. For instance, Fig. 3 shows the normalized thermal conductivity of SiNWs with different cross sectional geometries above room temperature. For all cross sectional geometries, the cross sectional area is the same of about 33 $nm^2$, which is in the universal region. Due to the intrinsically lowest SVR associated with rectangle, thermal conductivity of SiNW with rectangular cross section is the highest among all geometries, which is used as the normalization reference at each temperature, respectively. The triangular cross section has the highest SVR, leading to about 18% reduction of thermal conductivity compared to the rectangular cross section. In addition, this reduction ratio is robust above room temperature.

The threshold in cross sectional area, above which SVR can serve as a universal gauge for thermal conductivity, is found to be about 20 $nm^2$ in our study, which only corresponds to about 5 nm in diameter for cir-SiNW. This dimensional scale for the threshold is very close to the smallest diameter that current experimental fabrication can achieve [14]. Therefore, from practical point of view, SVR can be used as a universal gauge without any restriction of cross sectional area in experiment. However, from theoretical point of view, it is still of great interest to understand why the universal linear dependence of thermal conductivity on SVR breaks down below the threshold. The non-universal region differs from the universal region mainly in the following two aspects. First, thermal conductivity is less sensitive to SVR in non-universal region than that in universal region. Second, in the non-universal region, thermal conductivity depends not only on SVR, but also on the specific cross sectional geometry. Moreover, in this region, given the same SVR, tri-SiNW has a slightly higher thermal conductivity than that of the other two geometries.

To explore the underlying physical mechanism, we carry out a vibrational eigen-mode analysis by using lattice dynamics simulation. For a three-dimensional (3D) bulk material, under the harmonic approximation in lattice dynamics calculation, phonon eigen-modes can be calculated by diagonalizing the dynamical matrix which defines the interatomic interactions. By calculating all the eigen-modes in the first Brillouin zone, one can get the phonon dispersion relation. And the density of states DOS($\omega$) can be calculated by counting all the available eigen-modes within $\omega$ to $\omega+\Delta\omega$. The lattice dynamics calculation of a 1D nanowire is exactly analogous to that for the 3D bulk, except that the Brillouin zone is also now 1D, leading to dispersion only along the nanowire axis. In the calculation of nanowire, the cross section of the periodically repeated super cell should be larger than the actual cross section of the nanowire, which creates large enough vacuum space between neighboring nanowires to avoid spurious interactions. This technique has been routinely used in atomistic simulations of silicon nanowires by applying periodic boundary conditions [31-33]. In the present work, both the phonon eigen-modes and density of states are calculated by using lattice dynamics simulation package "General Utility Lattice Program" (GULP) [34], which is capable of modeling systems with all dimensionalities.

Phonon mode localization can be quantitatively characterized by the participation ratio $P_\lambda$ defined for each eigen-mode $\lambda$ as [35]

$$P_\lambda^{-1} = N \sum_i \left( \sum_\alpha \varepsilon_{i\alpha,\lambda}^* \varepsilon_{i\alpha,\lambda} \right)^2, \tag{1}$$

where $N$ is the total number of atoms, and $\varepsilon_{i\alpha,\lambda}$ is the $\alpha$th eigenvector component of eigen-mode $\lambda$ for the $i$th atom. It can provide detailed information about localization effect for each phonon mode in a given structure [8]. When macroscopically considering the influence of structure change on localization effect, averaged participation ratio ($P_{ave}$) defined as:

$$P_{ave} = \sum_{\omega} P(\omega)DOS(\omega) \qquad (2)$$

which is a very useful quantity to characterize the overall localization effect, with a smaller $P_{ave}$ indicating a stronger phonon localization effect [7]. The density of states $DOS(\omega)$ and eigen-modes of phonons are computed by using GULP in lattice dynamics calculations with periodic boundary condition. Fig. 4 shows the calculation result of $P_{ave}$ versus SVR. The blue and black lines draw the linear fit line of $P_{ave}$ in universal and non-universal region, respectively. In both universal and non-universal region, $P_{ave}$ decreases monotonically with the increase of SVR. In heat transport, the contribution to thermal conductivity mainly comes from the delocalized modes rather than the localized modes. The reduced $P_{ave}$ means that the percentage of delocalized mode decreases, leading to the reduction of thermal conductivity. Furthermore, the decreasing slope of $P_{ave}$ with respect to SVR is greater in the universal region than that in the non-universal region. This difference of the decreasing slope is the origin that thermal conductivity is less sensitive to SVR in non-universal region than that in universal region.

To further identify the important factor that causes the cross sectional geometry dependence of thermal conductivity in the non-universal region, we study the spatial distribution of the localized phonon modes. As demonstrated in Ref. [8], the spatial distribution of the localized phonon modes can be effectively visualized by the local energy $E_i$ defined as:

$$E_i = \sum_{\omega}\sum_{\lambda}\sum_{\alpha}(n+\frac{1}{2})\hbar\omega\varepsilon^*_{i\alpha,\lambda}\varepsilon_{i\alpha,\lambda}\delta(\omega-\omega_\lambda), \qquad (3)$$

where $n$ is the phonon occupation number given by the Bose-Einstein distribution, and $i$ denotes the $i$th atom that corresponds to a specific spatial location. To select the localized phonon modes, the summation in Eq. 3 only includes phonon modes with participation ratio less than 0.2.

Fig. 5 plots the normalized energy distribution at room temperature for localized phonon modes on the cross sectional (*YZ*) plane of SiNWs with SVR<0.14 (in the universal region). Due to the dangling bond atoms on the surface which break the otherwise perfect lattice periodicity, phonon modes are localized on the surface with mode amplitude exponentially decaying with the distance from the surface [36]. This is well depicted in Fig. 5 that the localized phonon modes reside close to the boundary of the cross sectional plane with only a few layers in thickness, which corresponds to the surface of SiNWs and can be accurately described by SVR. Moreover, since the corner atoms are obviously different from other surface atoms due to their lower coordination number, they can induce a stronger localization effect. This is indeed manifested in Fig. 5 by a higher local energy at corner atoms compared to that at other surface atoms. Similar enhancement of the local energy at the sharp corner is also found in the study of electronic properties of thin SiNWs [37, 38]. However, this difference between corner atoms and other surface atoms is not taken into account by SVR which can only measure the percentage of surface atoms in the whole system. On the other hand, as shown in Fig. 5, the weight of corner atoms in the surface atoms is quite small when the cross sectional area is large. Meanwhile, the local energy is only about 20% higher at the corner atoms than that at other surface atoms. Therefore, when the cross sectional area is large so that SVR is less than the threshold value of about 0.14, the influence of corner atoms can be neglected, and SVR can effectively describe the phonon localization effect on the surface, thus there is a universal dependence of thermal conductivity on SVR among different cross sectional geometries.

However, the situation is quite different when SVR>0.14 (in the non-universal region). Fig. 6 shows the normalized energy distribution of localized phonon modes for the small cross section case. In this case, the enhancement of the local energy at the corner persists, and the contrast in local energy between corner atoms and other surface atoms is much larger than that shown in Fig. 5. Furthermore, the ratio of corner atoms on the surface is enhanced a lot compared to the large cross section case. Due to these reasons, the influence of corner atoms can no longer be neglected at small cross sectional area. As a result, in addition to SVR, the corner atoms which have a stronger localization effect compared to other surface atoms, also play an important role in determining thermal conductivity. As shown in Fig. 6, tri-SiNW has only three corners, while rect-SiNW and cir-SiNW have four corners. Because of the fewer corner atoms, thermal conductivity of tri-SiNW is slightly higher than that of rect-SiNW and cir-SiNW in the non-universal region, given the same SVR. Thus the cross sectional geometry dependence of thermal conductivity in the non-universal region is caused by the stronger phonon localization effect at the corner atoms and their increasing significance in thin SiNWs.

Finally, we discuss the validity of other conventionally used gauge quantities for thermal conductivity, such as diameter and cross sectional area. Strictly speaking, the diameter is well defined only for circular cross section, but is ambiguous for other cross sectional geometries. Therefore, diameter is not suitable to serve as the universal gauge for thermal conductivity among different geometries. For the cross sectional area $A$, it can be defined for all geometries, and thermal conductivity usually increases monotonically with the increase of $A$ for any given cross sectional geometry. In principle, it can be used as the gauge for thermal conductivity of NWs with the same cross sectional geometry. But it is definitely not a universal gauge among different geometries, as it cannot take into account the specific surface morphology, which is a crucial factor in determining thermal conductivity at nanoscale. For example, we have demonstrated in Fig. 3 that even with the same cross section area, thermal conductivity of SiNWs with different geometries can differ by as large as 18% due to the intrinsically different SVR. Another example is that thermal conductivity of SiNWs with different cross sectional geometries can be quite close to each other due to the similar SVR, even if the cross sectional area differs by several times. For instance, room temperature thermal conductivity of rect-SiNW with $A$=62 nm$^2$ is very close to that of the tri-SiNW with $A$=213 nm$^2$ as shown in the inset of Fig. 2(a), while the cross sectional area differs by more than a factor of 3. Therefore, SVR as the gauge for thermal conductivity has its advantage over other conventionally used gauge quantities in the sense that it is a universal one among different geometries, when the cross sectional area is greater than the threshold.

## 3. Conclusions

In summary, we have studied thermal conductivity of SiNWs with different cross sectional geometries and cross sectional area up to 806 nm$^2$. A universal linear dependence of thermal conductivity on surface-to-volume ratio is found for SiNWs with modest cross sectional area larger than about 20 nm$^2$, regardless of the specific cross sectional geometry. Moreover, we found the absolute value of the slope for the linear fit line depends on both temperature and length, due to the temperature and length dependence of thermal conductivity of SiNWs. In addition, with the same cross sectional area (larger than 20 nm$^2$), tri-SiNW has the lowest thermal conductivity due to its intrinsically highest SVR, which might be favorable in thermoelectric applications. Our study provides not only a universal gauge for thermal conductivity among different cross sectional geometries, but also a simple approach to tune thermal conductivity based on geometry consideration.

## Acknowledgements

Jie Chen thanks Nuo Yang for helpful discussions. The work has been supported by an NUS grant, R-144-000-285-666. The work has been also supported in part by a grant from the Asian Office of Aerospace R&D of the US Air Force (AOARD-114018). GZ was supported by the Ministry of Science and Technology of China (Grant Nos. 2011CB933001).

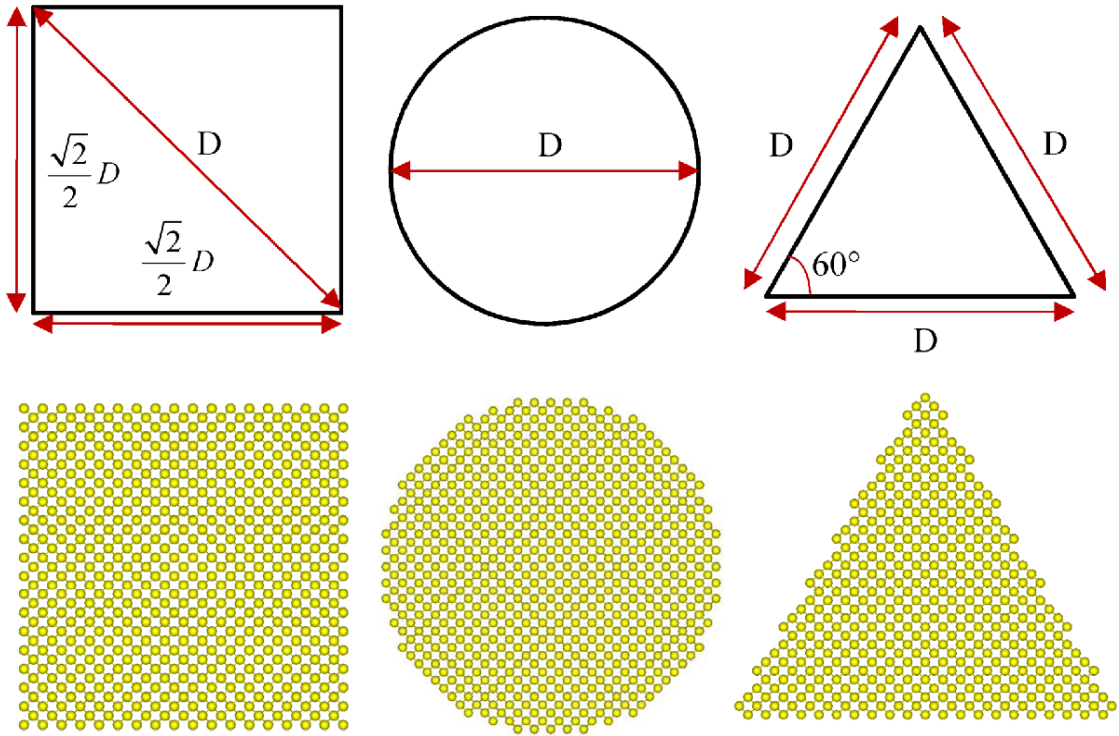

Figure 1. Cross sectional view of the ideal geometry and actual lattice structure of SiNWs. The diameter *D* is defined for all geometries as the maximum distance between the surface atoms in the cross sectional plane.

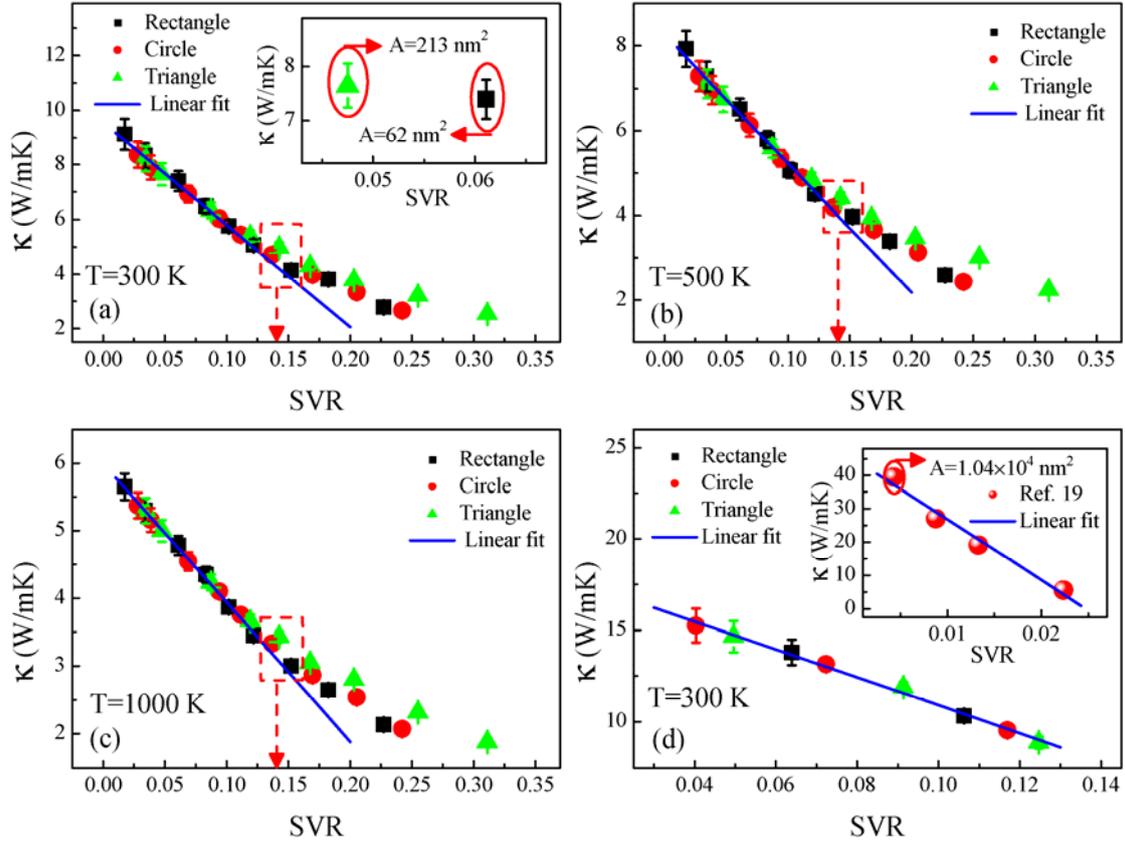

Figure 2. Thermal conductivity ($\kappa$) of [100] SiNWs versus surface-to-volume ratio (SVR) for different temperatures: (a) 300 K, (b) 500 K, and (c) 1000 K. The NW length is fixed at $L$=3.26 nm. The black square, red circle and green triangle denote thermal conductivity of rect-SiNWs, cir-SiNWs and tri-SiNWs, respectively. (d) Thermal conductivity of SiNWs for different cross sectional geometries at 300 K. The NW length is $L$=6.52 nm. The inset in (d) shows experimental results at 300 K from Ref. 19. In all figures, the blue lines are the best-fitting ones, and $A$ is the cross sectional area. The absolute value of the slope in each figure is (a) slope=37.5±1.2, (b) slope=29.2±0.8, (c) slope=20.6±0.6, and (d) slope=76.2±2.3. The red dash lines draw the threshold region of about SVR=0.14, which corresponds to a cross sectional area of about 10 nm$^2$, 15 nm$^2$, and 23 nm$^2$, respectively, for rectangular, circular, and triangular cross section. SVR in this figure is defined as the ratio of the number of surface atoms to the total number of atoms.

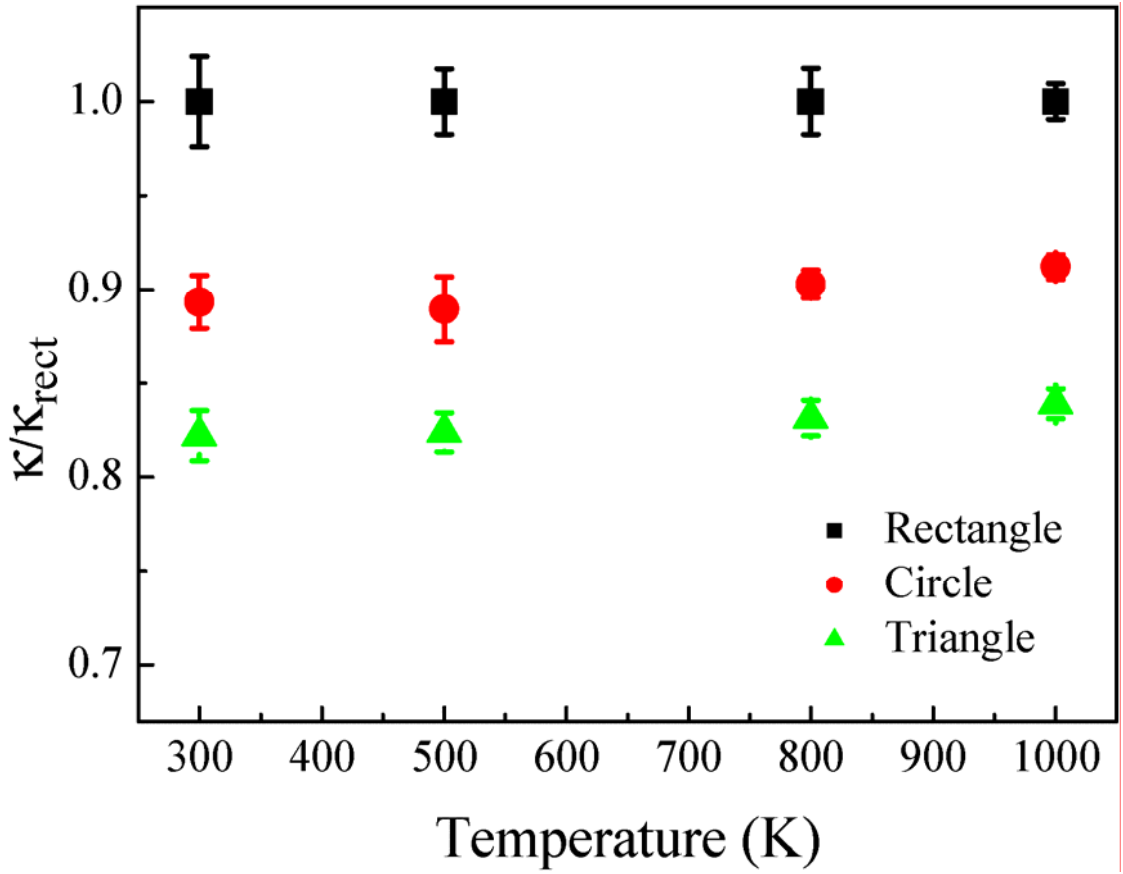

Figure 3. Normalized thermal conductivity of SiNWs versus temperature for different cross sectional geometries. Thermal conductivity of rect-SiNWs at each temperature is used as reference. The black square, red circle and green triangle denote thermal conductivity of rect-SiNWs, cir-SiNWs and tri-SiNWs, respectively. The cross sectional area is the same (about 33 nm$^2$) for all cross sectional geometries, and the nanowire length is 3.26 nm.

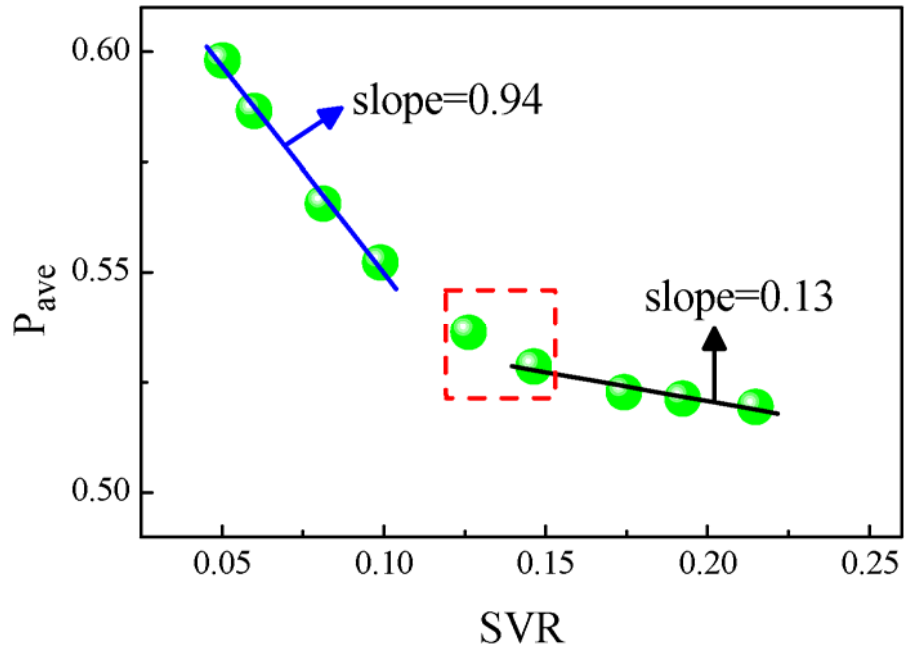

Figure 4. Averaged participation ratio ($P_{ave}$) versus SVR. The blue and black lines are the best-fitting ones. The red dash line highlights the threshold region. All eigen-modes are computed by using GULP based on rect-SiNWs along [100] direction, and the length of supercell is kept as one unit cell for different cross sectional areas in lattice dynamics calculations with periodic boundary condition.

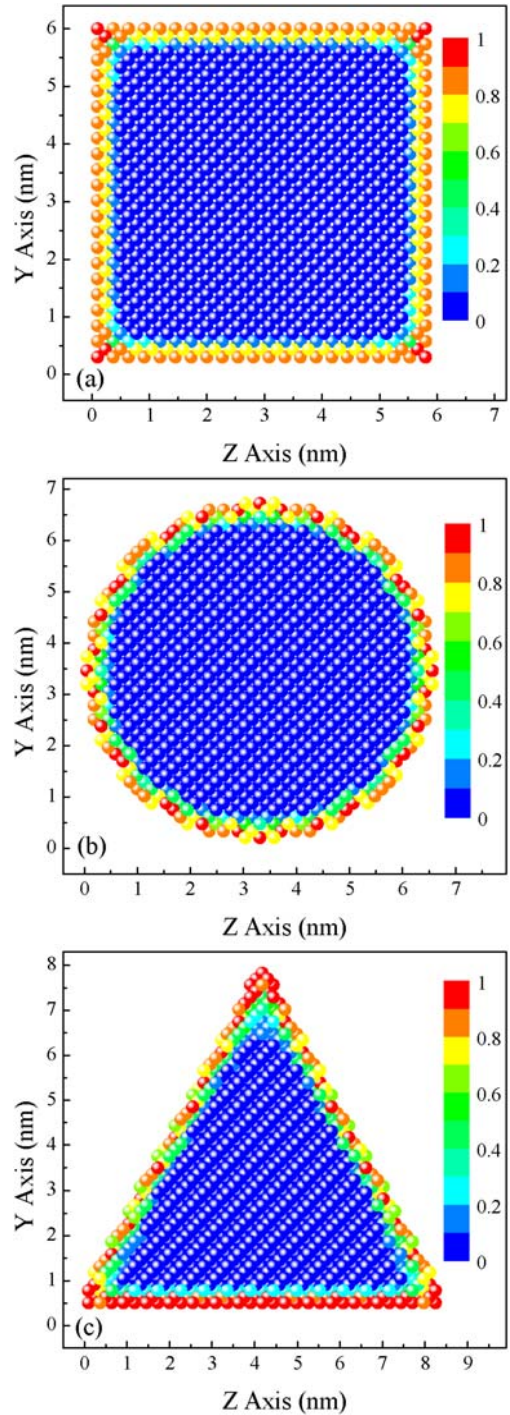

Figure 5. Normalized energy distribution at 300 K for the localized phonon modes on the cross sectional (*YZ*) plane of SiNWs with large cross sectional area. Positions of the circles denote the different locations on the *YZ* plane, and intensity of the energy is depicted according to the colour. (a) Rect-SiNW with *A*=32.5 nm$^2$; (b) Cir-SiNW with *A*=33.4 nm$^2$; (c) Tri-SiNW with *A*=28.8 nm$^2$.

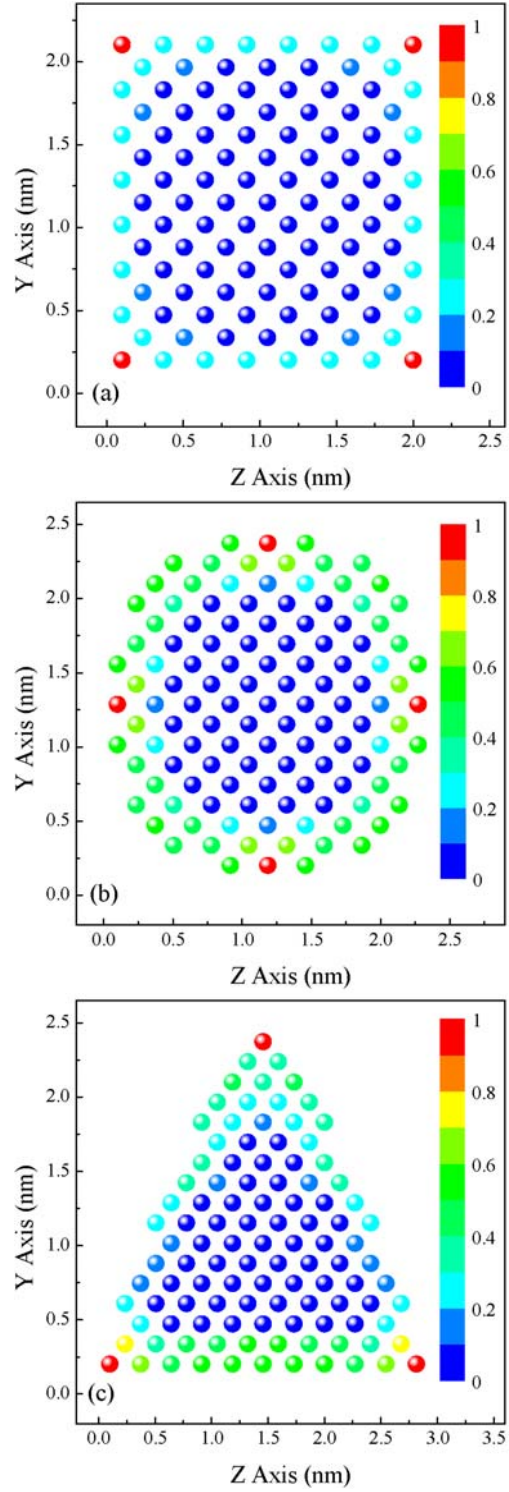

Figure 6. Normalized energy distribution at 300 K for the localized phonon modes on the cross sectional plane of SiNWs with small cross sectional area. Positions of the circles denote the different locations on the plane, and intensity of the energy is depicted according to the colour. (a) Rect-SiNW with $A$=3.6 nm$^2$; (b) Cir-SiNW with $A$=3.7 nm$^2$; (c) Tri-SiNW with $A$=3.2 nm$^2$.